\documentclass{article}
\usepackage{cite,fullpage}
\usepackage{amsmath,amssymb,amsfonts}
\usepackage{orcidlink}
\usepackage{cite}
\usepackage{algorithmic}
\usepackage{textcomp}
\usepackage{xcolor}
\def\BibTeX{{\rm B\kern-.05em{\sc i\kern-.025em b}\kern-.08em
    T\kern-.1667em\lower.7ex\hbox{E}\kern-.125emX}}
\usepackage{lipsum} 
\usepackage[margin=1in]{geometry}
\usepackage{amsmath,amssymb,amsthm,amsfonts,array,graphicx,multirow,pdflscape,mathrsfs}
\hypersetup{breaklinks=true, colorlinks=true, linkcolor={blue!90!black}, citecolor={blue!90!black}, urlcolor={blue!90!black}}

\usepackage{dcolumn}
\usepackage{bm}
\usepackage[mathlines]{lineno}
\usepackage{adjustbox}
\usepackage{url}
\usepackage{braket}
\usepackage{listings}
\usepackage{caption}
\usepackage{booktabs}
\usepackage{multirow}
\usepackage{siunitx}
\usepackage{algorithm}

\usepackage{caption}

\definecolor{ForestGreen}{RGB}{34,139,34}

\usepackage[normalem]{ulem} %

\usepackage[algo2e,linesnumbered,vlined,algoruled]{algorithm2e}

\usepackage{marginnote}

\usepackage{thmtools}
\usepackage{nameref}
\usepackage{doi,cite}
\usepackage[nameinlink,noabbrev]{cleveref}

\newcommand{\orbv}{\mathcal{O}_v}
\newcommand{\amc}{\tilde{A}}
\newcommand{\orbe}{\mathcal{O}_e}
\usepackage{xspace}
\newcommand{\qaoa}{\textsf{QAOA}\xspace}
\newcommand{\maqaoa}{\textsf{ma-QAOA}\xspace}
\newcommand{\bestsymqaoa}{\textsf{best-1sym-QAOA}\xspace}
\newcommand{\maxsymqaoa}{\textsf{max-sym-QAOA}\xspace}
\newcommand{\randsymqaoa}{\textsf{rand-group-QAOA}\xspace}

\DeclareRobustCommand{\abbrevcrefs}{%
\Crefname{theorem}{Thm.}{Thms.}%
\Crefname{figure}{Fig.}{Figs.}%
\Crefname{algorithm}{Alg.}{Algs.}%
\Crefname{section}{Sec.}{Secs.}%
}

\newcommand{\sCref}[1]{{\abbrevcrefs\Cref{#1}}}

\usepackage{authblk}
\begin{document}

\title{Multi-Angle QAOA Does Not Always Need All Its Angles}

\author[a,b]{Kaiyan Shi}

\author[c]{Rebekah Herrman}

\author[d]{Ruslan Shaydulin}
\author[d]{Shouvanik Chakrabarti}
\author[d]{Marco Pistoia}

\author[a]{Jeffrey Larson}

\affil[a]{Mathematics and Computer Science Division, Argonne National Laboratory, {\tt jmlarson@anl.gov}}
\affil[b]{Dept.~of Computer Science, University of Maryland, {\tt kshi12@umd.edu}}
\affil[c]{Industrial and Systems Engineering, University of Tennessee at Knoxville}
\affil[d]{Global Technology Applied Research, JPMorgan Chase}
\date{}                     %
\renewcommand\Affilfont{\itshape\small}

\maketitle

\begin{abstract}
Introducing additional tunable parameters to quantum circuits is a powerful way of improving performance without increasing hardware requirements. A recently introduced multiangle extension of the quantum approximate optimization algorithm (\maqaoa) significantly improves the solution quality compared with \textsf{QAOA} by allowing the parameters for each term in the Hamiltonian to vary independently. Prior results suggest, however, considerable redundancy in  parameters, the removal of which would reduce the cost of parameter optimization. In this work we show numerically the connection between the problem symmetries  and the parameter redundancy by demonstrating that symmetries can be used to reduce the number of parameters used by \maqaoa  without decreasing the solution quality. We study Max-Cut on all 7,565 %
connected, non-isomorphic 8-node graphs with a nontrivial symmetry group and show numerically that in 67.4\% %
of these graphs, symmetry can be used to %
reduce the number of parameters with no decrease in the objective, with the average ratio of parameters reduced by $\textbf{28.1\%}$. %
Moreover, we show that in 35.9\% of the graphs %
this reduction can be achieved by simply using the largest symmetry. %
For the graphs where reducing the number of parameters leads to a decrease in the objective, the largest symmetry can be used to reduce the parameter count by $\textbf{37.1\%}$ at the cost of only a $\textbf{6.1\%}$ decrease in the objective. We demonstrate the central role of symmetries by showing that a random parameter reduction strategy leads to much worse performance. %
\end{abstract}

\section{Introduction}

Quantum hardware has improved rapidly in recent years~\cite{arute2019quantum,wu2021strong,madsen2022quantum,ringbauer2022universal}, opening up the possibility of demonstrating quantum advantage on a relevant practical problem. Combinatorial optimization problems are commonly considered targets for near-term quantum devices~\cite{mcgeoch2013experimental, mohseni2022ising}, with the quantum approximate optimization algorithm (\qaoa)~\cite{Hogg2000,farhi2014quantum} as a promising candidate algorithm because of its low hardware resource requirements~\cite{harrigan2021quantum,2204.05852,2206.06290,Shaydulin2021EM}.  %

\qaoa solves optimization problems using a parameterized circuit composed of layers of alternating operators, with two operators being  evolutions with a Hamiltonian encoding the objective function and a problem-instance-independent mixer Hamiltonian. %
The evolution times are free parameters (often called angles), which are optimized with the goal of maximizing the expected quality of the measurement outcomes. 
The success of variational quantum algorithms with a large number of trainable parameters such as quantum neural networks and the variational quantum eigensolver~\cite{Cerezo2021} motivated the introduction of additional parameters in QAOA. Intuitively, adding additional parameters to the algorithm based on the structure of the problem %
can only increase the circuit expressiveness and thereby can only improve the algorithm's performance.  %

Multiangle \qaoa (\maqaoa) is a modification of \qaoa that incorporates additional parameters~\cite{Herrman2021} by allowing the parameter associated with each term in the problem and mixer Hamiltonian to vary independently.  %
 \maqaoa  has been shown to solve Max-Cut on star graphs exactly using only one layer, whereas \qaoa achieves an approximation ratio of only 0.75. The improvement in the quality of the solution achieved by the introduction of the parameters is modest, however, suggesting that the large number of parameters does not translate to a highly expressive circuit. Moreover, preliminary \maqaoa  research has shown that parameters tend to cluster around multiples of $0.25\pi$~\cite{herrman2022investigating}. Together, these observations suggest that
the number of parameters in \maqaoa  can be reduced without affecting the solution quality. %

In this work we demonstrate the connection between the redundancy in \maqaoa parameters and the problem symmetries. %
Specifically, we reduce the number of parameters by setting the parameters connected by a chosen symmetry to be equal. We consider the problem of Max-Cut and show numerically that on $68.0\%$ of graphs that have a nontrivial symmetry group, the number of parameters can be reduced on average by $28.1\%$ by using one of the symmetries without decreasing the objective function value.
 Inspired by this observation, we propose a modification of \maqaoa  that uses the full symmetry group (\maxsymqaoa). The full symmetry group can be obtained efficiently for many classes of graphs, and fast heuristic solvers can be used in practice~\cite{shaydulin2021classical}. We show that \maxsymqaoa reduces the number of parameters by $37.1\%$ at the cost of only a $6.1\%$ decrease in the objective.
 Moreover, we provide evidence of the centrality of the symmetries by showing that a random strategy with the same number of parameters yields much worse performance. 
 
 This paper is organized as follows. First, we introduce binary optimization, \qaoa, and graph symmetry background material in \sCref{sec:background}. In \sCref{sec:methods} we then discuss the methods used in this work. In \sCref{sec:results} we discuss our results, and we conclude with a discussion in \sCref{sec:discussion}.

\section{Background} \label{sec:background}

We first briefly review the relevant background material and introduce the notation. 

\subsection{Binary Optimization Problem}\label{subsec:binarybackground}

We consider binary optimization(BO) problems of the form $\max_{x \in \{0,1\}^n} f(x),$ where $f(x)$ is a non-negative objective function over the Boolean cube $\{0,1\}^n$. It is often a sum of other functions that describe the system, called clauses. 

When solving BO problems on quantum hardware, we construct a cost Hamiltonian $H_c$ that encodes $f(x)$, so that $H_c\ket{x} = f(x)\ket{x}.$ Then the optimization problem becomes $$\max_{x \in \{0,1\}^n} \bra{x}H_c\ket{x}.$$

The outcome of the algorithm is marked as $x^*$, and algorithm performance is typically quantified by the approximation ratio $r \in [0,1]$ given by
\begin{align} \label{eqn:def_r}
r := \frac{f(x^*)}{\max f(x)} = \frac{\bra{x^*}H_c\ket{x^*}}{\max \bra{x}H_c\ket{x}}.
\end{align}

\subsection{\qaoa}\label{subsec:qaoabackground}

\qaoa is a hybrid quantum-classical algorithm that finds approximate solutions to combinatorial optimization problems~\cite{farhi2014quantum}.
To solve a given optimization problem with \qaoa, one must construct a cost Hamiltonian $H_c$ that encodes the objective function and a mixer Hamiltonian $H_m$. Let $U(\gamma, C) = e^{-i H_c \gamma}$ and $U(\beta, B) = e^{-i H_m \beta}$, where $\gamma$ and $\beta$ are free parameters.
These two unitaries are applied to an initial state $\ket{s}$, which is an eigenvector of $H_m$. The outcome of $p$ iterations of the algorithm is denoted $\ket{\vec{\gamma}, \vec{\beta}}_p$ and is

\begin{align*}
    \ket{\vec{\gamma}, \vec{\beta}}_p = &U(\beta_p, H_m) U(\gamma_p,H_c) \ldots \\
    & U(\beta_1, H_m) U(\gamma_1, H_c)\ket{s}.
\end{align*}
The parameters $\gamma$ and $\beta$ are chosen to maximize
\begin{equation*}
E(\vec{\gamma}, \vec{\beta}) = \bra{\vec{\gamma}, \vec{\beta}} H_c \ket{\vec{\gamma}, \vec{\beta}}.
\end{equation*}
Measuring the state $\ket{\vec{\gamma},\vec{\beta}}$ gives an approximate solution to the BO problem encoded by $H_c$. 

\maqaoa  is similar to \qaoa; however, the definitions of $U(\gamma, C)$ and $U(\beta, B)$ are changed to

\begin{equation*}
    U(\vec{\gamma}, C) = e^{-i \sum_a C_a\gamma_a}
\end{equation*}
and
\begin{equation*}
    U(\vec{\beta}, B) = e^{-i \sum_b X_b\beta_b},
\end{equation*}
where $C_a$ is a clause in the objective function and $X_b$ is the Pauli-x operator acting on qubit $b$. Throughout this work, $r_{\textsf{y}}$ refers to the approximation of \textsf{y-QAOA}, where \textsf{y} is a variation of \qaoa.

\subsection{\qaoa on Max-Cut Problem}\label{subsec:maxcutbackground}
The Max-Cut problem is well studied in \qaoa literature (e.g.~\cite{harrigan2021quantum,shaydulin2019evaluating}) and is thus a natural problem to consider when studying \qaoa variants. Given a simple graph $G=(V,E)$, the Max-Cut problem aims to partition $V$ into two disjoint sets so that the number of edges with endpoints in both sets is maximized. This problem is NP-hard to solve exactly. %

When solving the Max-Cut problem using \qaoa, the cost Hamiltonian is 
\begin{equation*}
H_c = \sum_{uv \in E}\frac{1}{2} (I-Z_{u}Z_{v}),
\end{equation*}
and the mixer Hamiltonian is typically 
\begin{equation*}
H_m = \sum_{v \in V} X_v,
\end{equation*}
where $Z_{v}$ and $X_v$ are single-qubit Pauli operators acting on qubit $v$. Each layer of the \qaoa circuit has one $\gamma$ for all edges and one $\beta$ for all vertices. The number of tunable parameters is just $2p$, independent of the graph size. 

In $\maqaoa$, each vertex and each edge has its own angle. Thus, there are $(|E|+|V|) \cdot p$ parameters to optimize in this modified algorithm. One drawback to this approach is that finding $(|E|+|V|) \cdot p$ parameters can be difficult, especially as the size of the problems grows. However, the algorithm has better performance than \qaoa has on average~\cite{Herrman2021}. %

\subsection{Graph Symmetries} \label{subsec:symmetrybackground}

A graph automorphism is a permutation of the vertex set of a graph, $\sigma: V\rightarrow V$, that satisfies the condition that a pair of vertices, $(u,v)$, forms an edge if and only if the pair $(\sigma(u),\sigma(v))$ also forms an edge. Automorphisms can be represented by products of disjoint cycles, $\sigma = \pi_1 \ldots \pi_k$, where $\pi_l = (i_1, i_2, \ldots, i_j)$ and each entry in the cycle is a unique integer. In this notation, $\sigma(i_a) = i_{a+1}$ modulo $j$. %
Any set of automorphisms can generate a corresponding vertex and edge orbit, which are the equivalence classes of the vertices (edges) of a graph $G$ under the action of the automorphism.

A generator of a group is a set of automorphisms $(\sigma_1, \ldots, \sigma_n)$ containing group elements so that (possibly, repeated) application of the generators on themselves and each other can produce all elements in the group. In this work, we call the group generator of the automorphism group of the graph $G$ the symmetry generator, and it can generate corresponding vertex and edge orbits. In this paper we denote the vertex (edge) orbit of the symmetry generator as the \emph{maximum vertex (edge) orbit}, written as $\orbv (\orbe)$. %

\section{Methods}\label{sec:methods}

We use the symmetry structure of the problem to reduce the number of parameters in \maqaoa{}. Symmetry is known to impact \qaoa performance~\cite{shaydulin2021classical}. To analyze the role of symmetries, we introduce three modifications called \bestsymqaoa, \maxsymqaoa, and \randsymqaoa.

 \textsf{sym-QAOA} selects a single automorphism of the target graph and assigns the same angle to all vertices in the same vertex orbit and the same angle to all edges in the same edge orbit. It then optimizes the \qaoa parameters and executes the resulting \qaoa circuit. \bestsymqaoa runs \textsf{sym-QAOA} over all automorphisms of $G$ and selects the one that gives the largest Max-Cut value.
 
When using \maxsymqaoa to solve Max-Cut on a graph $G$, the first step is to compute the symmetry generator of $G$ and determine the corresponding maximum vertex orbit, $\orbv$, and edge orbit, $\orbe$. The algorithm requires $|\orbv| + |\orbe|$ parameters, where each element in the same vertex orbit or edge orbit receives the same parameter. The algorithm samples $|\orbv| + |\orbe|$ parameters randomly as initial parameters and runs the \qaoa variational quantum circuit as a subroutine using those parameters. \qaoa optimization steps are applied until the solution converges to the optimal solution or until the desired number of iterations has been performed. The formal algorithm is described in \sCref{alg:max-sym-QAOA}.

Finding the generating set of the automorphism group of a graph is an extra step in \maxsymqaoa compared with \maqaoa. The time complexity of this step is at most quasi-polynomial since its polynomial time equivalent graph isomorphism problem can be solved by a quasi-polynomial algorithm~\cite{babai2016graph}. Also, many polynomial time heuristics exist for specific classes of graphs
such as \textbf{nauty}~\cite{mckay2007nauty, mckay2013nauty} used in this work. %

\randsymqaoa groups vertices in the problem graph randomly into $|\orbv| $ sets and edges randomly into $|\orbe|$ sets, so that the number of parameters is the same as that of \maxsymqaoa.

The generator used in \maxsymqaoa is usually a set of automorphisms, so \maxsymqaoa is not always contained in \bestsymqaoa, which ranges over all single automorphisms. Thus, \maxsymqaoa may perform better than \bestsymqaoa.

\begin{algorithm2e}[t]
    \SetKwInOut{Input}{Input}
    \SetKwInOut{Output}{Output}
    \caption{\maxsymqaoa\label{alg:max-sym-QAOA}}
    
		\Input{Graph $G$, number of layers $p$.}
		\Output{Optimized $\{\vec{\beta}, \vec{\gamma}\}$ approx. max-cut $\amc$}
		
		Construct cost Hamiltonian $H_c$ from $G$
		
		Find the symmetry generator of $G$ and corresponding maximum vertex/edge orbit sets $\orbv$, $\orbe$
	
		Sample $|\orbv| + |\orbe|$ initial parameters $\{\vec{\beta}, \vec{\gamma}\} = \{\beta_0, \ldots, \beta_{|\orbv-1|}, \gamma_0, \ldots, \gamma_{|\orbe-1|}\}$. Fix vertices/edges in the same orbit to have the same value 
	    $\ket{\psi(\vec{\beta},\vec{\gamma})} \gets \textsf{QAOAcirc}(\{\vec{\beta}, \vec{\gamma}\}, p)$
	    
        $E(\vec{\beta}, \vec{\gamma}) \gets \bra{\psi(\vec{\beta},\vec{\gamma})} H_c \ket{\psi(\vec{\beta},\vec{\gamma})}$ 
        
        $\{\vec{\beta}, \vec{\gamma}\}, x^* \gets$ classical optimization algorithms to optimize $E(\vec{\beta}, \vec{\gamma})$ 
\end{algorithm2e}

\section{Results}\label{sec:results}
In this work we implement one iteration of \maxsymqaoa, \randsymqaoa, and \bestsymqaoa,  using the graph descriptions from~\cite{mckay}. We then compare the algorithms with one another and \maqaoa  using the data found in~\cite{lotshawdataset}. %
We use COBYLA to optimize the \qaoa parameters, although we expect similar results to be obtained with other gradient-free and gradient-based local methods.

\subsection{\bestsymqaoa vs. \maqaoa}

\begin{figure}[t]
    \centering
    \includegraphics[width=0.42\textwidth]{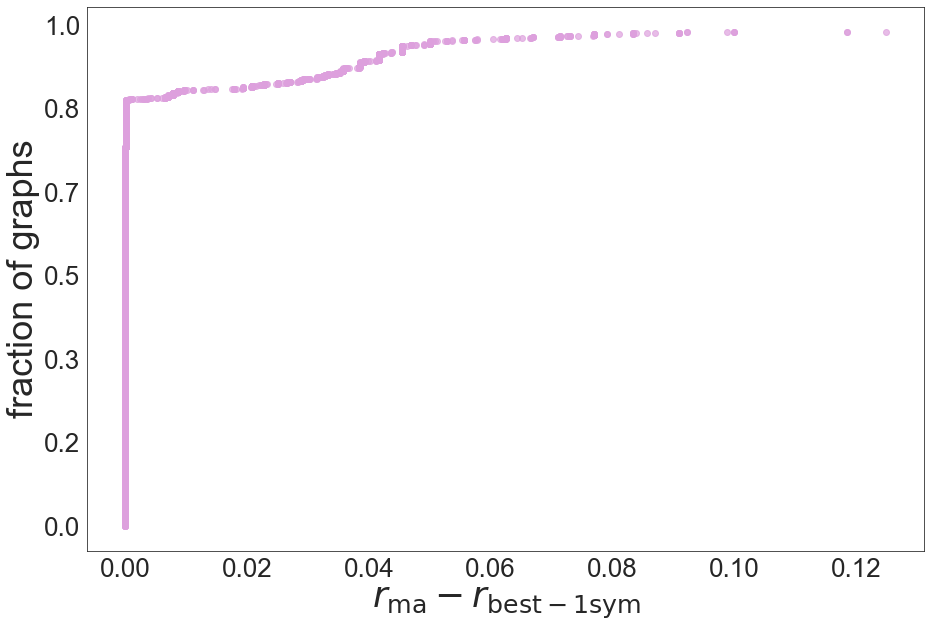}
    \caption{Difference of approximation ratio $r$ between \maqaoa  and \bestsymqaoa, as defined in \eqref{eqn:def_r}. For most graphs with symmetry, one symmetry can be used to reduce the number of parameters without affecting the solution quality.\label{fig:ma_best_sym}}
\end{figure}

Among all $7,565$ graphs with nontrivial symmetries, \bestsymqaoa has fewer parameters than \maqaoa has on $5,918$ graphs. Thus, we analyze only \bestsymqaoa on those graphs. \Cref{fig:ma_best_sym} shows the difference in approximation ratios between \bestsymqaoa and \maqaoa  for these $5,918$ graphs. \bestsymqaoa has the same approximation ratio as \maqaoa has  on $5,097$ graphs, which is approximately $86.1\%$ of the studied graphs, while using on average $28.1 \%$ fewer parameters than \maqaoa uses. %

We also quantify the ratio of the difference in the Max-Cut (approximation ratio) values of \maqaoa and \bestsymqaoa to the difference in the Max-Cut (approximation ratio) values of \maqaoa and \qaoa as 
\begin{align}\label{eq:k_best_1sym}
    k_{\bestsymqaoa} := \frac{f(x^*_{\textsf{ma}}) - f(x^*_{\bestsymqaoa})}{f(x^*_{\textsf{ma}}) - f(x^*_{\qaoa})},
\end{align}
where $f(x^*_{\textsf{y}})$ denotes the approximate Max-Cut found by $\textsf{y}-QAOA$. When $k=0$, \bestsymqaoa recovers \maqaoa; and when $k=1$, \bestsymqaoa performs the same as \qaoa. Thus, this ratio indicates whether \bestsymqaoa performance is closer to \maqaoa  performance or \qaoa performance. In this study, the denominator of the ratio is always nonzero.

It is encouraging that the approximation ratios for \bestsymqaoa and \maqaoa are similar. 
But this finding is of limited value if both methods use the same number of parameters.
We therefore consider the quantity $l_{\bestsymqaoa}$, which is the relative difference in parameters between 
\maqaoa and \bestsymqaoa (as compared with the difference in the number of parameters between \maqaoa and \qaoa).
That is, 
\begin{align}\label{eq:l_best_1sym}
    l_{\bestsymqaoa} :=  \frac{|E|+|V| - (|\mathcal{O}_e(\sigma)| + |\mathcal{O}_v(\sigma)|)}{|E|+|V|-2},
\end{align} 
where $|\mathcal{O}_v(\sigma)|$ and $|\mathcal{O}_e(\sigma)|$ are the number of vertex orbits and edge orbits, respectively, induced by the automorphism $\sigma$. This ratio determines how close the number of parameters in \bestsymqaoa is to either \maqaoa  or \qaoa, depending on whether the ratio is closer to $0$ or not.

\subsection{\maxsymqaoa vs. \maqaoa }

\begin{figure}[t]
    \centering
    \includegraphics[width=0.42\textwidth]{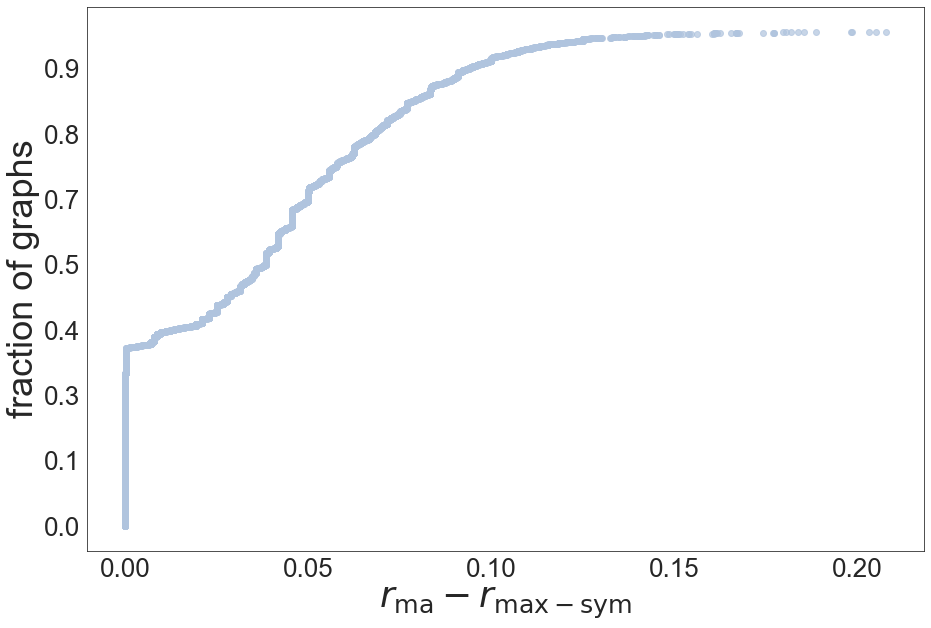}
    \caption{Difference of approximation ratio $r$ between \maqaoa  and \maxsymqaoa, as defined in \eqref{eqn:def_r}. In a plurality of graphs with symmetry, simply using the largest symmetry leads to a reduction in the number of parameters with no impact on solution quality.
    \label{fig:ma_max_sym}}
\end{figure}

Of the 11,117 connected, non-isomorphic 8-vertex graphs, $3,552$ have only trivial symmetries. In these cases \maxsymqaoa is \maqaoa. Thus, our analysis focuses on the $7,565$ graphs that contain nontrivial symmetry, where \textsf{max-sym-QAOA} has fewer parameters to optimize over. As shown in \sCref{fig:ma_max_sym}, \maxsymqaoa performs as well as \maqaoa  on 2,713 of these graphs. Furthermore, it performs the same as \qaoa on 30 graphs, which is only about $0.4\%$ of the graphs with nontrivial symmetry. These results indicate that \maxsymqaoa performance is comparable, even though it requires fewer parameters. 

We define the ratio of the difference in the Max-Cut values of \maqaoa  and \maxsymqaoa to the difference in Max-Cut values of \maqaoa and \qaoa as
\begin{align}\label{eq:k_max_sym}
    k_{\textsf{max-sym}} := \frac{f(x^*_{\textsf{ma}}) - f(x^*_{\textsf{max-sym}})}{f(x^*_{\textsf{ma}}) - f(x^*_{\qaoa})}.
\end{align}
When $k=0$, \maxsymqaoa recovers \maqaoa; and when $k=1$, \maxsymqaoa performs the same as \qaoa. Thus, this ratio indicates whether \maxsymqaoa's performance is closer to \maqaoa's  performance or \qaoa's performance. In this study, the denominator of the ratio is always nonzero.

We quantify the ratio of the reduction of parameters from \maqaoa  to \maxsymqaoa over the difference in parameters between \maqaoa  and \qaoa as
\begin{align}\label{eq:l_max_sym}
    l_{\textsf{max-sym}} :=  \frac{|E|+|V| - (|\orbe| + |\orbv|)}{|E|+|V|-2},
\end{align}
since \maqaoa  uses $|E|+|V|$ parameters, \maxsymqaoa requires $|\orbe| + |\orbv|$ parameters, and \qaoa uses two parameters. %

\begin{figure}
    \centering
    \includegraphics[width=0.42 \textwidth]{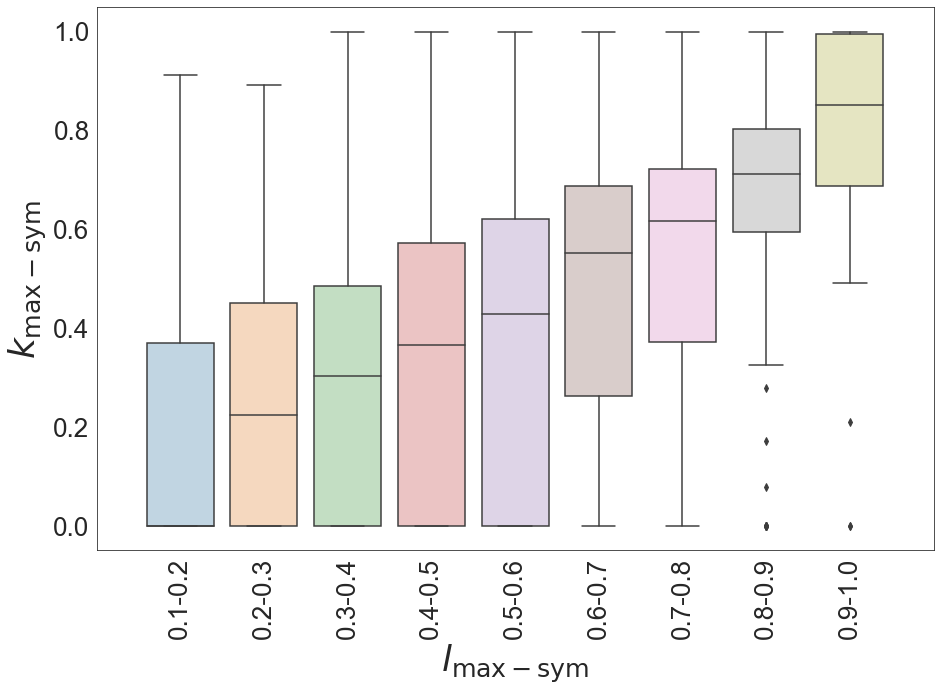}
    \caption{Comparison of $k_{\textsf{max-sym}}$ and $l_{\textsf{max-sym}}$ for \maxsymqaoa, as defined in \eqref{eq:k_max_sym} and \eqref{eq:l_max_sym}. There is a positive correlation between these two variables. \label{fig:params vs. maxcut}}
\end{figure}

\Cref{fig:params vs. maxcut} shows that a positive correlation exists between the quantities $k$ and $l$. Among results that achieve the result equivalent to %
\maqaoa, the number of parameters reduced is spread nearly evenly over $[0.1,0.8]$. Since there are almost no points between $[0,0.1]$, \maxsymqaoa almost always requires at least $10\%$ fewer parameters than \maqaoa requires. $l_{\textsf{max-sym}}$ averaged over all those graphs with nontrivial symmetries is $0.37$.

Note that specifically for those graphs where reducing the number of parameters leads to a decrease in the objective, \textsf{max-sym-QAOA} can reduce the parameter count by $37.1\%$ at the cost of only a $6.1\%$ decrease in the objective. 
\subsection{Evidence for Central Role of Symmetries}

\begin{figure}[b!]
    \centering
    \includegraphics[width=0.42\textwidth]{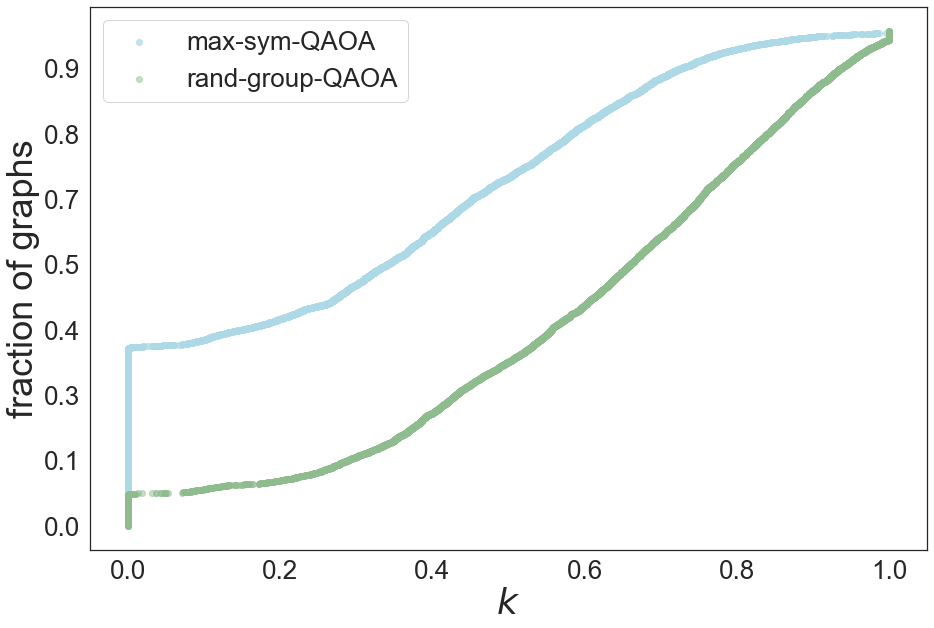}
    \caption{Fraction of graphs achieving ratio $k$ for \maxsymqaoa and \textsf{rand-group-QAOA}. If a random parameter reduction is used, the performance deteriorates significantly, suggesting the central role of symmetries.
    \label{fig:max_vs_rand}}
\end{figure}

\maxsymqaoa is also compared with \randsymqaoa, which groups vertices and edges randomly so that the corresponding number of parameters is the same as that of \maxsymqaoa. \Cref{fig:max_vs_rand} demonstrates the centrality of symmetries to \maqaoa parameter redundancy by showing that the parameter reduction strategy of \maxsymqaoa has a clear advantage over \randsymqaoa in terms of solution quality. %

\subsection{\maxsymqaoa vs. \bestsymqaoa}

In this section we compare \maxsymqaoa and \bestsymqaoa on the 5,918 graphs for which \text{best-sym-QAOA} requires fewer parameters than does \maqaoa. %

\begin{figure}[t]
    \centering
    \includegraphics[width=0.42\textwidth]{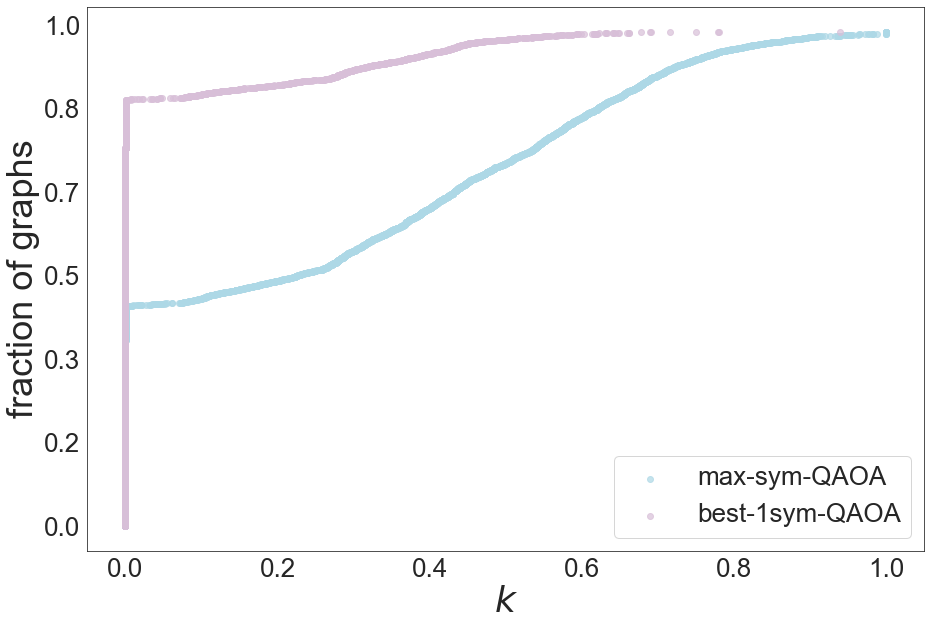}
    \caption{Fraction of graphs achieving ratio $k$ for \maxsymqaoa and \bestsymqaoa, as defined in \eqref{eq:k_best_1sym} and \eqref{eq:k_max_sym}. \label{fig:max_vs_best}}
\end{figure}

\Cref{fig:max_vs_best} indicates that \bestsymqaoa has $k=0$ on nearly twice as many graphs as \maxsymqaoa. Additionally, $k<0.6$ for the majority of graphs solved with \bestsymqaoa while $k$ is spread over $[0,1]$ with \maxsymqaoa.

 Although \maxsymqaoa does not perform as well as \bestsymqaoa on the majority of graphs, the average $l_{\textsf{max-sym}}$ is around $0.39$ while the average $l_{\textsf{best-sym}}$ is only $0.31$, so \bestsymqaoa has more parameters than \maxsymqaoa, on average.

\section{Discussion}\label{sec:discussion}

In this work we demonstrate the connection between the parameter redundancy in \maqaoa and the symmetries of the problem to be optimized. Specifically, we show that the number of parameters in \maqaoa can often be dramatically reduced without affecting the solution quality. To that end, we introduce three \qaoa variations that require fewer parameters than \maqaoa: \bestsymqaoa, \maxsymqaoa, and \randsymqaoa. The three algorithms assign classical parameters based on the symmetries (automorphisms) of the underlying problem graph. We evaluate these algorithms on all connected, non-isomorphic 8-vertex graphs and compare the results with those of \maqaoa.

In most cases,  \maxsymqaoa requires at least $10\%$ fewer parameters than \maqaoa does, while maintaining a comparable approximation ratio, which is the primary metric of \qaoa success. In fact, in over one-third of the connected 8-vertex graphs with nontrivial symmetry, \maxsymqaoa finds the same approximate Max-Cut as does \maqaoa. Furthermore,  a positive correlation exists between the number of parameters reduced and the reduction in the approximation ratio, as expected. Additionally, significantly more graphs have $k=0$ when solved with \maxsymqaoa than \randsymqaoa, implying that \maxsymqaoa outperforms \randsymqaoa in general. On the other hand, significantly more graphs have $k=0$ when solved with \bestsymqaoa than \maxsymqaoa, yet \maxsymqaoa needs fewer parameters (on average) than does \bestsymqaoa.

Thus, out of \bestsymqaoa, \maxsymqaoa, and \randsymqaoa, \randsymqaoa appears to have the worst performance while \bestsymqaoa appears to have the closest performance to \maqaoa on these small graphs. The failure of \randsymqaoa demonstrates the importance of symmetry to parameter setting in \qaoa. %

Approximately one-third of the graphs considered in this study  had only trivial symmetry, and \maxsymqaoa is equivalent to \maqaoa in these cases. Nonetheless, numerical evidence in ~\cite{Herrman2021} suggests that for these graphs the redundancy in parameters is also present. Therefore, an interesting future direction is understanding how the redundancy in parameters can be reduced for graphs with no symmetries. %

Sauvage et al.~proposed using symmetries to improve the performance of variational quantum algorithms~\cite{sauvage2022building}. They observed that in \maqaoa applied to the Max-Cut problem, the number of parameters can be reduced and the trainability improved by using an approach equivalent to \maxsymqaoa described in this work. Unlike \cite{sauvage2022building}, we highlight the role of symmetries in quantum optimization by showing that symmetry-based parameter reduction leads to much better performance than does a random approach. Moreover, we consider utilizing a part of the symmetry group (\bestsymqaoa).

\section*{Acknowledgments}
This work was supported by the U.S.~Department of Energy, Office of Science, Office of Advanced Scientific Computing Research, Accelerated Research for Quantum Computing program.
Herrman acknowledges the NSF award CCF-2210063 and the DARPA ONISQ program under award W911NF-20-2-0051.

\bibliographystyle{IEEEtran}
\bibliography{references}

\vfill

\section*{Disclaimer}

This paper was prepared for information purposes with contributions from the Global Technology Applied Research center of JPMorgan Chase. This paper is not a product of the Research Department of JPMorgan Chase or its affiliates. Neither JPMorgan Chase nor any of its affiliates make any explicit or implied representation or warranty and none of them accept any liability in connection with this paper, including, but not limited to, the completeness, accuracy, reliability of information contained herein and the potential legal, compliance, tax or accounting effects thereof. This document is not intended as investment research or investment advice, or a recommendation, offer or solicitation for the purchase or sale of any security, financial instrument, financial product or service, or to be used in any way for evaluating the merits of participating in any transaction.

\vfill

\framebox{\parbox{.90\linewidth}{\scriptsize 
The submitted manuscript has been created with contributions from
        UChicago Argonne, LLC, Operator of Argonne National Laboratory (``Argonne'').
        Argonne, a U.S.\ Department of Energy Office of Science laboratory, is operated
        under Contract No.\ DE-AC02-06CH11357.  The U.S.\ Government retains for itself,
        and others acting on its behalf, a paid-up nonexclusive, irrevocable worldwide
        license in said article to reproduce, prepare derivative works, distribute
        copies to the public, and perform publicly and display publicly, by or on
        behalf of the Government.  The Department of Energy will provide public access
        to these results of federally sponsored research in accordance with the DOE
        Public Access Plan \url{http://energy.gov/downloads/doe-public-access-plan}.}}

\end{document}